\documentclass[11pt]{amsart}

\usepackage[margin=1.2in]{geometry}
\usepackage{times}

\usepackage{amsmath}
\usepackage{amssymb}
\usepackage{amsfonts}
\usepackage{amsthm}
\usepackage{mathrsfs}
\usepackage[all]{xy}
\usepackage{endnotes} 
\usepackage{cite}
\usepackage{graphicx}

\begin{document}

\begin{titlepage}

\begin{center}
\LARGE{\textbf{Spatially resolved edge currents and guided-wave electronic states in graphene}}\\[1.5 cm]
\large{M. T. Allen$^{1}$, O. Shtanko$^{2}$, I. C. Fulga$^{3}$, A. Akhmerov$^{4}$, K. Watanabi$^{5}$, T. Taniguchi$^{5}$,  P. Jarillo-Herrero$^{2}$, L. S. Levitov$^{2}$, and A. Yacoby$^{1*}$}\\[1.0 cm]

$^{1}$ Department of Physics, Harvard University. Cambridge, MA 02138.

$^{2}$ Department of Physics, Massachusetts Institute of Technology. Cambridge, MA 02139.

$^{3}$ Department of Condensed Matter Physics, Weizmann Institute of Science. Rehovot, Israel.

$^{4}$ Kavli Institute of Nanoscience. Delft University of Technology. Lorentzweg 1. 2628 CJ Delft. The Netherlands. 

$^{5}$ Environment and Energy Materials Division, National Institute for Materials Science. 1-1 Namiki, Tsukuba, Ibaraki, 305-0044 Japan.\\[1.0 cm]

\textbf{$^{*}$Corresponding author. E-mail: yacoby@physics.harvard.edu}
\end{center}

\end{titlepage}




\begin{quote}
\small{\textbf{Abstract:}  
A far-reaching goal of graphene research is exploiting the unique properties of carriers to realize extreme nonclassical electronic transport. Of particular interest is harnessing wavelike carriers to guide and direct them on submicron scales, similar to light in optical fibers. Such modes, while long anticipated,  
have never been demonstrated experimentally. In order to explore  this behavior, we employ superconducting interferometry in a graphene Josephson junction to reconstruct the real-space supercurrent density using Fourier methods. Our measurements reveal charge flow guided along crystal boundaries close to charge neutrality.
We interpret the observed edge currents in terms of guided-wave states, confined to the edge by band bending and transmitted as plane waves. 
As a direct analog of refraction-based confinement of light in optical fibers, such nonclassical states afford new means for information transduction and processing at the nanoscale.}\\
\end{quote}

Electrons in Dirac materials such as graphene can be manipulated using external fields to control electron refraction and transmission in the same way that
optical interfaces in mirrors and lenses can manipulate light
~\cite{RMP1, Klein1, Klein5, Shytov1}.  
Several of the key ingredients, including phase-coherent Klein transmission and reflection~\cite{Klein2, Klein4, Klein3}, ballistic transport~\cite{Mayorov1} and transverse focusing on micrometer scales~\cite{MagneticFocusing1}, have already been established. 
One promising yet unexplored direction, which we investigate here, is the quasi-1D confinement of electrons in direct analogy to refraction-based confinement of photons in optical fibers. Electronic guided modes formed by a line gate potential, while discussed in the literature
\cite{GuidedModes1,GuidedModes2,GuidedModes3,Klein7}, have so far evaded direct  experimental detection. Extending the fiber optics techniques to the electronic domain
is key for achieving control of electron waves at a level comparable to that for light in optical communication systems.\\

Rather than pursuing the schemes 
discussed in Refs.\cite{GuidedModes1,GuidedModes2,GuidedModes3,Klein7}, here we explore modes at the graphene edges.
The atomically sharp graphene edges provide a natural vehicle for band bending near the boundary which then confines the electronic waves in the direction transverse to the edge. The resulting guided ``fiber-optic'' modes propagate along the edge as plane waves, decaying into the graphene bulk as evanescent waves. 
In analogy with optical fibers, 
such modes are situated outside the Dirac continuum (see Fig. 1A,B). 
As discussed below, the mode frequency in monolayer graphene is
\begin{equation}\label{eq:w=uk}
\omega=\tilde v |k|-{\textstyle \frac{i}2} \gamma(k),\quad |\tilde v|<v=10^6\,{\rm m/s}
,
\end{equation}
where the damping $\gamma(k)$ accounts for scattering by disorder. For the practically interesting regime of disorder originating from edge roughness, the damping is expected to quickly vanish at long electron wavelengths near charge neutrality, scaling as $\gamma(k)\sim k^2$. In this regime, as discussed below, mode scattering by the edge is near-specular and is thus immune to backscattering. 
This approach to carrier guiding is particularly appealing because of the ease with which band bending at the graphene edge can be realized,
as well as because 
there is no threshold for fiber-optic states to occur:
they are induced 
by an edge potential of either sign, positive or negative, no matter how weak (see discussion below and in the Supplementary Methods). 
The presence of such guided modes 
enhances the density of current-carrying states at the 
edge for doping near charge neutrality, while uniform behavior is recovered at higher carrier concentrations (see Fig. 1C and Fig. S1). \\

The edge currents associated with guided states, anticipated at zero magnetic field, have so far eluded experimental detection due to the challenge of imaging current with submicron spatial resolution. In particular, scanning tunneling spectroscopy (STS)  images density of states but not current flow~\cite{Ritter1, Crommie1}, while macroscopic conductivity cannot distinguish the edge and bulk contributions~\cite{Morpurgo2, Dominik1}. Additionally, these techniques covered disparate length scales, with STS probing atomic-scale lengths and transport covering micron to millimeter distances.  High resolution density of states measurements along unzipped carbon nanotube boundaries and graphene quantum dots were obtained spectroscopically using STS, but 
the dispersive nature and current-carrying capacity of these states was unresolved~\cite{Ritter1, Crommie1}. \\

With this motivation, we developed a technique to spatially image electric current pathways and applied it to high-mobility graphene. We employ Fraunhofer interferometry in a graphene Josephson junction to reconstruct the spatial structure of the electronic states which transmit supercurrent.
To implement this approach, we measure gated Josephson junctions consisting of graphene coupled to superconducting titanium/aluminum electrodes (Fig. 1D).  A gate electrode is used to tune the carrier density $n$. In order to access the intrinsic properties of graphene at densities near charge neutrality, we isolate the flakes from substrate-induced disorder through placement on thin hexagonal boron nitride (hBN) substrates~\cite{Dean1}.  We study four bilayer and one monolayer devices, all of which exhibit similar behavior (Table S1).   Figures 1E-H 
exemplify transport data from one of the bilayer devices described above.  Upon sweeping DC current bias $I_{DC}$, a sharp transition in resistance between dissipationless and normal metal behavior appears at a critical current $I_c$, a signature of the Josephson effect (Fig. 1E,F).\\

We employ superconducting quantum interference to extract a spatially resolved image of the supercurrent density across the flake.  It is possible to obtain real-space information because application of a magnetic flux $\Phi$ through the junction area induces a position-dependent superconducting phase difference $\Delta \phi(x)= 2\pi \Phi x/ \Phi_0 W$ parallel to the graphene/contact interface~\cite{TinkhamBook}, where $\Phi_0=h/2e$ is the flux quantum, $h$ is Planck's constant, $e$ is the elementary charge, and $W$ is the width of the flake (Fig. 1D).  The resulting interference is displayed in Figure 1E, a plot of differential resistance $dV/dI$ as a function of $I_{DC}$ and magnetic field $B$.   Measurements of the AC voltage drop $dV$ across the junction in response to an AC current modulation $dI$ were conducted using lockin techniques in a dilution refrigerator at 10 mK, well below the critical temperature of Al.  The critical current $I_c$, obtained by extracting the value of $I_{DC}$ at the maximum derivative $dV/dI$, exhibits modulations in $B$ field that arise from Fraunhofer-like magnetic interference. As the flux threading the junction winds the superconducting phase along the length of the contact, the critical current $I_c$ can be expressed quantitatively as the magnitude of the complex Fourier transform of the current density distribution $J(x)$.  That is, $I_c=|\mathcal{I}_c(B)|$, where
\begin{equation}
\mathcal{I}_c(B)= \int_{-W/2}^{W/2} J(x)\cdot e^{2\pi iLBx/\Phi_0} dx 
\end{equation}
where $L$ is the distance between contacts.  Relevant
for wide junctions ($L\ll W$) such that the current is only a function of one coordinate, Equation (2) provides a simple and concise description of our system.   The spatial distribution of supercurrent thus dictates the shape of the interference pattern~\cite{TinkhamBook, Dynes1, DasSarma1}.  \\

Our results, obtained with this technique, show a strikingly different behavior at high and low carrier densities.  We observe conventional  Josephson behavior with uniform current flow at high density, for which the normalized critical current
$I_c(B)/I_c(0) \sim |\sin (\pi \Phi/ \Phi_0) / (\pi \Phi / \Phi_0) |$
 is described by single-slit Fraunhofer diffraction  (Fig. 1E). Defining features of such interference include a central lobe of width $2\Phi_0$ and side lobes with period $\Phi_0$ and decaying $1/B$ amplitude.  However, near the Dirac point, our results exhibit a striking departure from this picture and show 
an enhancd ``SQUID-like'' interference (Fig. 1F) ~\cite{SQUID1}. Such behavior arises when supercurrent is confined to edge channels and is characterized by slowly decaying sinusoidal oscillations of period $\Phi_0$.  Importantly, these two regimes are easily distinguishable without much analysis by observing the width of the central lobe which is twice as wide for the uniform case as compared to the case of edge flow.  \\

The real-space current distribution can be quantitatively obtained by inverting 
the relation in Eq.(1) with the help of the Fourier techniques of Dynes and Fulton~\cite{Dynes1} (see Supplementary Methods). A more numerically expensive Bayesian estimation of the current distribution produces current distributions and standard error estimates that agree with the Fourier techniques (see Supplementary Methods and Fig. S2).
The resulting electron density map reveals strong confinement of supercurrent to the edges of the crystal near the Dirac point (Fig. 1H).  This phenomena is a robust experimental feature seen in all five devices.  The width of the edge currents, extracted quantitatively from Gaussian fits, is on the order of the electron wavelength ($\sim 200$ nm) and roughly consistent across multiple samples.  This value is likely an upper bound because the peak width is manifested in the decay envelope of the interference pattern, external factors that suppress the critical current amplitude at high $B$, such as activation and decay of the Al superconducting gap, may also contribute to peak broadening.  Although the maximum fields of $B \sim 10-25$ mT are a small fraction of the critical field of Al ($H_c\sim 100$ mT),  
the actual edge states may be narrower.  At high electron density, conventional single-slit Fraunhofer interference is recovered (Fig. 1E), suggesting a uniform distribution of supercurrent (Fig. 1G).\\
 
By tuning carrier concentration with a gate electrode, our measurements reveal coexistance of edge and bulk modes at intermediate densities in (Fig. 2, monolayer graphene).  Starting at the Dirac point, the device exhibits SQUID-like quantum interference through charge neutrality and the spatial image of supercurrent reveals edge-dominated transport (Fig. 2A-C).  As density is increased, bulk current flow increases, 
crossing over to mostly uniform flow (Fig. 2D) and conventional Fraunhofer interference at high electron density (Fig. 2E).  To track the evolution of edge and bulk currents with density, we plot line cuts of the individual contributions in Figure 2E, where the gate voltage corresponding to the charge neutrality point is identified as a dip in the current amplitude.  Notably, raw interference near the Dirac point (Fig. 2E) and at high electron concentration (Fig. 2F) exhibit the salient features that distinguish edge-dominated from bulk-dominated transport, including a width of $\phi_0$ versus $2\phi_0$ of the central lobe, as well as Gaussian versus $1/B$ decay of the lobe amplitudes for low and high densities, respectively.  Finally, due to the spin and momentum conserving nature of Andreev reflection, the observed supercurrents suggest an absence of strong spin and valley scattering at the edge in high quality samples. \\

Similarly, we systematically explore the correspondence between edge and bulk flow in bilayer graphene and detect boundary currents in the presence of broken crystal inversion symmetry (Fig. 3).  As the Fermi energy approaches the Dirac point from the hole side, the bulk is suppressed faster than the edge, leading to emergence of robust edge currents near zero carrier density. Upon tuning the Fermi level to high electron doping, a uniform distribution re-emerges (Fig. 3A,B). We note that the range in hole density over which the bulk
contribution is recovered varies in different devices.  At hole doping, an intrinsic p-n junction emerges at the graphene-electrode boundary due to contact-induced charge transfer, a phenomenon intensively studied using both transport and photocurrent techniques~\cite{Blake1, Kern1, Wu1}. In this regime, we find that edge transmission is augmented relative to the bulk, perhaps because an intrinsic gap at the p-n interface can block bulk flow more effectively in bilayer graphene. (In this device, current distributions are not plotted at the immediate Dirac point due to suppression of proximity-induced superconductivity at high normal state resistances.)  Furthermore, we apply an interlayer electric field $E$ to break crystal inversion symmetry and induce a bandgap~\cite{Gap1, Gap2}, manifested as a gate-tunable insulating state at the Dirac point (Fig. 3C,D).  In this regime, we find that conductance is mediated by edge currents that enclose the bulk, even in the presence of a field-induced gap (Fig. 3E,F).\\

Our measurements establish that edge currents dominate electronic transport in graphene at zero magnetic field near the Dirac point.  As demonstrated below, such edge currents can arise from `fiber-optic' electronic states formed in a line potential  localized near the edge. 
As a simple model exhibiting an analogy between electronic guided states and refraction-based confinement of light in fiber optics, we consider massless Dirac particles in graphene monolayer
in the presence of a line potential:
\begin{equation}
\epsilon\psi(x,y)=\left( v\boldsymbol{\sigma} {\boldsymbol p}+V(x)\right)\psi(x,y)
,\quad v\approx 10^6\,{\rm m/s}
.
\end{equation}
We seek plane-wave solutions of the form $\psi(x,y)=e^{iky}\phi(x)$. Here $k$ is the wavevector component along the line and $\phi(x)$ is a two-component spinor wavefunction depending on the transverse coordinate, and $\sigma_i$ are Pauli matrices. This problem 
can be tackled by a gauge transformation 
\begin{equation}
\phi(x)=e^{-i\theta(x)\sigma_x}\tilde\phi(x)
,\quad
\theta(x)=\frac1{\hbar v}\int_0^x V(x')dx'
.
\end{equation} 
Such a transformation  generates a mass term in the Dirac equation and eliminates the potential $V(x)$  entirely. 
We further simplify the resulting equation using the identity $e^{i\theta\sigma_x}\sigma_y e^{-i\theta\sigma_x}=\sigma_y \cos(2\theta)-\sigma_z \sin(2\theta)$ to obtain
$
\epsilon\tilde\phi(x)=\hbar v(-i\sigma_x\partial_x+k\sigma_y \cos(2\theta(y))-k\sigma_z \sin(2\theta(x))
\tilde\phi(x)
$.\\

As a simple example, we consider the case of an armchair edge, for which the problem on a half plane for carriers in valleys $K$ and $K'$ is equivalent to the problem on a full plane for a single valley. Applying the above equation to a potential  localized in an interval $-d<x<d$ and focusing on the long-wavelength modes such that $kd\ll1$, we can approximate 
$\theta(x)\approx \frac12 u\,{\rm sgn\,}(x)$ with the parameter $u=\frac1{\hbar v}\int_{-d}^{d} V(x')dx'$. 
We arrive at the seminal Jackiw-Rebbi problem for a Dirac equation with a mass kink
\begin{equation}
\epsilon\tilde\phi(x)=\hbar v (-i\sigma_x\partial_x+\sigma_y \tilde k+\sigma_z m(x))\tilde\phi(x)
,\quad \tilde k=k\cos u, \ m(x)=-k\sin u\,{\rm sgn\,}(x)
.
\end{equation}
The Jackiw-Rebbi problem can be solved exactly and explicitly~\cite{Jackiw1}. Guided-wave states for such a Hamiltonian are described as products of the zero-mode state found for $\tilde k=0$ and the plane wave factors $e^{iky}$. The energies of these states are simply $\epsilon=\pm\hbar \tilde v k$ with the sign given by ${\rm sgn\,} ( m(0+)-m(0-))$. This gives the dispersion in Eq.(\ref{eq:w=uk}) with $\tilde v=\pm v \cos u$ and the sign ${\rm sgn\,}(\sin u)$. 
Since $|\tilde v|<v$, for each $k$ the energies of these states lie outside the bulk continuum $|\epsilon|>\hbar v|k|$, which ensures decoupling from the bulk states and confinement to the region near the $x=0$ line. 
The connection with the theory of zero modes indicates the robustness of such confinement. Similar fiber-optic states are obtained for an edge potential 
in graphene bilayer (see Fig. 1B). \\

To assess compatibility with this model, we compare supercurrent density measurements with 
a theoretical prediction of density of states (Fig. 4).  Real space line cuts of current flow $J(x)$ in bilayer device $BL3$ at fixed densities are provided in Fig. 4a, showing edge currents near the Dirac point and a continuous evolution of bulk flow.  Theoretical plots of density of states as a function of position (Fig. 4B), obtained from the `fiber-optics' model of transport, exhibit qualitatively similar behavior.  For the simulation, an effective delta function potential approximation is used with the best-fit value $\lambda = 0.5 $ eV$\cdot$nm (see Supplementary Methods).\\

Despite the edge roughness inevitable in our devices, the observed edge currents are robust.  
This is consistent with the guided-wave model. Indeed, the long-wavelength guided modes are characterized by a large transverse confinement lengthscale, since
the evanescent wave decay defines a transverse confinement 
lengthscale on the order of electron wavelength $\lambda$, which is much larger than the edge roughness scale $\xi$. In this case, most of the mode wavefunction resides far from the edge, at distances $r\sim\lambda \gg \xi$, and is thus not susceptible to disorder scattering. For typical wavelength values $\lambda\sim 10-100$ nm, which exceeds the atomic scale characteristic for the edge disorder by two to three orders of magnitude, we expect backscattering suppression by as much as $(\lambda/\xi)^2\sim 10^4-10^5$ times to be readily achievable [see Eq.(\ref{eq:w=uk}) and Supplementary Methods]. This is similar to optical guided states in so-called ``low-contrast'' optical fibers, which feature a similar suppression of disorder scattering by a large `Thomson factor'   $\lambda / \xi$. This suppression is key for achieving extremely long mean free paths in such fibers.\\

To the best of our knowledge, the fiber-optic modes is the only model which is fully consistent with the observations. For instance, edge density accumulation can influence the edge potential and in principle also support guided edge currents.  However, the fact that the charge neutrality points for both edge and bulk roughly coincide in density $n$ suggests an absence of a positional charge imbalance  on a large scale (Fig. 2F). 
 In addition, edge-dominated current flow is observed near the Dirac point but not at higher densities, 
the behavior not expected for strong edge doping.  Arguments against random density inhomogeneities 
 (of the kind familiar from our previous studies of electron-hole puddles)
include the reproducibility of edge currents with width on the order of electron wavelength across many samples, as well as the observation that edge currents tend to 
be stronger in clean samples with ballistic
Fabry-P\'erot interference.  Large charge inhomogeneities across the sample would 
suppress Fabry-P\'erot interference and are thus unlikely. \\

One appealing aspect of the fiber-optic model is that it can naturally accommodate 
a wide range of different microscopic physical mechanisms discussed theoretically in the literature~\cite{Bandbending1, Jackiw1, Zigzag1, Akhmerov1, Guinea1} that may produce an edge potential.  Examples include pinning of the Fermi energy to the low-energy states due to broken A/B sublattice symmetry~\cite{Zigzag1, Akhmerov1, Guinea1}, density accumulation caused by dangling bonds or trapped charges at the boundaries, or electrostatics~\cite{Bandbending1, Jackiw1}. The competition of these effects can produce a complex dependence of the edge potential $V(x)$ on carrier density. Pinpointing the precise microscopic origins of the edge potential requires further study. \\

Lastly, it is widely known that the A/B sublattice imbalance for broken bonds at the edge
can lead to edge modes in pristine graphene at neutrality.  Such ``dispersing zero-mode states'' 
can exist even in the absence of  a line potential, forming 
edge modes for 
an atomically perfect zigzag edge~\cite{Zigzag1, Akhmerov1, Guinea1}. However, 
our simulations for disordered edge show that these states are highly localized on the disorder length scale, 
and also that edge roughness induces strong scattering between the states at the boundary and in the bulk, producing relatively short mean free path values and hindering ballistic propagation. 
We therefore conclude that such states are unlikely to contribute to the observed boundary currents.  \\

To summarize, we present evidence of zero field edge currents in a graphene Josephson junction.  These findings underscore the relevance of edge effects 
for graphene transport and provide a new tool for exploration of boundary currents, a topic of emerging interest.  
Formed due to band bending at the edge,
the observed 
guided states demonstrate confinement of electron waves at a level comparable to that for light in photonic systems.
This defines a new mode for transmission of electronic signals at the nanoscale.  We also anticipate this work will inspire more detailed investigations of boundary states in graphene and other materials.  Our techniques open the door to fast spatial imaging of current distributions along more complicated networks of domains in larger crystals, increasingly relevant as graphene sheets are scaled to larger dimensions~\cite{Muller1}. Such capabilities are also of fundamental interest due to the predicted topological nature of edge states along stacking domain boundaries in bilayer graphene~\cite{Mele1, Alden1}.\\

\section*{Acknowledgments}
The authors thank O. Dial, B. Halperin, V. Manucharyan, and J. Sau for helpful discussions. This work is supported by the Center for Integrated Quantum Materials (CIQM) under NSF award 1231319 (LSL and OS) and the U.S. DOE Office of Basic Energy Sciences, Division of Materials Sciences and Engineering under award DE-SC0001819 (PJH, MTA, AY). Nanofabrication was performed at the Harvard Center for Nanoscale Systems (CNS), a member of the National Nanotechnology Infrastructure Network (NNIN) supported by NSF award ECS-0335765.  AA was supported by the Foundation for Fundamental Research on
Matter (FOM), the Netherlands Organization for Scientific Research
(NWO/OCW). ICF was supported by the European Research Council under
the European Union's Seventh Framework Programme (FP7/2007-2013) / ERC
Project MUNATOP, the US-Israel Binational Science Foundation, and the
Minerva Foundation for support.

\section*{Figure legends}

\textbf{Fig. 1.}${}\quad$\textbf{`Fiber-optic' modes and spatially resolved current imaging in a graphene Josephson junction}.  \textbf{(A, B)} Guided edge modes induced by an intrinsic band bending near crystal boundary, for single-layer and bilayer graphene (schematic). Mode frequencies positioned outside the Dirac continuum ensure mode decoupling from the bulk states. Guided modes exist for any edge potential no matter how weak. In a single layer, mode velocity changes sign as the potential strength increases, see Eq.(5). In a bilayer, the modes occur in pairs [\textit{green and red curves}: dispersion for positive and negative potential strength, respectively].
\textbf{(C)} The 
guided modes are manifested through peaks in 
the density of current-carrying states at the crystal boundaries, prominent near charge neutrality
(\textit{red}: $n=0.05\times10^{11} cm^{-2}$; \textit{blue}: $n=2.5\times10^{11} cm^{-2}$).  \textbf{(D)} Schematics of superconducting interferometry in a graphene Josephson junction, 
which
is used to image the spatial structure of current-carrying states. A flux is threaded through the junction area to produce interference patterns, as current bias $V_{sd}$ is applied through
the superconducting electrodes and the voltage drop across the device is recorded.  Carrier density $n$ is tuned by a gate voltage $V_{\rm b}$.  \textbf{(E, F)} 
The recorded interference pattern is of a single-slit Fraunhofer type at high carrier density, turning into a SQUID-like interference near neutrality
(colorscale is $dV/dI\,(\Omega)$ for device $BL1$). \textbf{(G, H)} Current flow, extracted from the interference data using Fourier techniques, is uniform at high carrier density and 
peaks at the crystal edges for carrier density close to neutrality.   \\

\textbf{Fig. 2.}${} \quad$\textbf{Gate-tunable evolution of edge and bulk current-carrying states in graphene}. \textbf{(A)} Edge-dominated 
SQUID-like interference pattern at neutrality
in device $ML1$ ($n=2.38\times 10^9$ cm$^{-2}$; colorscale is $dV/dI (\Omega)$).  \textbf{(B, C)} Real-space image of current flow confined to the boundaries over a range of densities near neutrality, shown alongside with the raw interference data
(corresponding to the white box in (D)).  
\textbf{(D)} A real-space map of current flow as a function of electron concentration reveals coexistence of edge and bulk modes at intermediate densities.  \textbf{(E)} Conventional Fraunhofer pattern for uniform current flow at high electron density ($n=7\times 10^{11}$ cm$^{-2}$).  \textbf{(F)} Comparison of current amplitudes along the edge (red) and bulk (blue) from the plot in panel (C).
Current flow is edge-dominated near neutrality. Note that 
minima for both contributions coincide in $n$, indicating that a positional 
edge/bulk density offset is not present.\\

\textbf{Fig. 3.}${} \quad$\textbf{Boundary currents in bilayer graphene in the presence of broken crystal inversion symmetry}. \textbf{(A)} Spatially resolved supercurrent map in device $BL2$, in a normalized plot of $J(x)/J_{max}(x)$.  Edge-dominated transport occurs near charge neutrality, while an increasing bulk contribution is tuned with carrier concentration.  \textbf{(B)} Comparison of current amplitudes along the edge (red) and through the bulk (blue) from panel (A).  Enhanced edge currents
are prominent at neutrality, whereas a uniformly distributed flow is recovered at high densities.  The normal state conductance $G(e^2/h)$ 
{\it vs.} carrier density is also shown (black).   \textbf{(C)} Measurement schematic for superconducting interferometry in a dual-gated bilayer graphene Josephson junction.  
A dual-gated device consists of bilayer graphene flake on hBN with a suspended top gate, where application of voltages $V_{\rm t}$ and $V_{\rm b}$ on the top and back gates enables independent control of the transverse electric field $E$ and carrier density $n$.  \textbf{(D)} Resistance map as a function of $V_{\rm b}$ and $V_{\rm t}$ for bilayer $BL4$.  Enhanced resistance at high $E$ fields indicates the emergence of a gate-tunable insulating state due to broken crystal inversion symmetry. \textbf{(E)} Spatially-resolved boundary currents as a function of $E$ field.  The vertical axis is a trace along the red path labeled in (B). \textbf{(F)} Sequence of Fraunhofer measurements at various locations on the current map in panel (E).\\

\textbf{Fig. 4.}${} \quad$\textbf{`Fiber-optics' theoretical model of transport in graphene}.  \textbf{(A)} Real-space maps of measured current flow $J(x)$ in bilayer device $BL3$ at fixed carrier densities on the hole side, showing edge currents near the Dirac point and a continuous evolution towards
bulk flow.  \textbf{(B)} Theoretical plot of spatially resolved density of states in bilayer graphene at fixed carrier densities for edge waveguide model. For the simulation, an effective delta function potential approximation is used with the best-fit value $\lambda = 0.5 $ eV$\cdot$nm (see SOM).
Band mass of bilayer graphene is taken $0.04 m_e$ where $m_e$ is electron mass.\\

\newpage
\textbf{Figure 1}
\begin{figure}[!h]
\includegraphics[width=160mm]{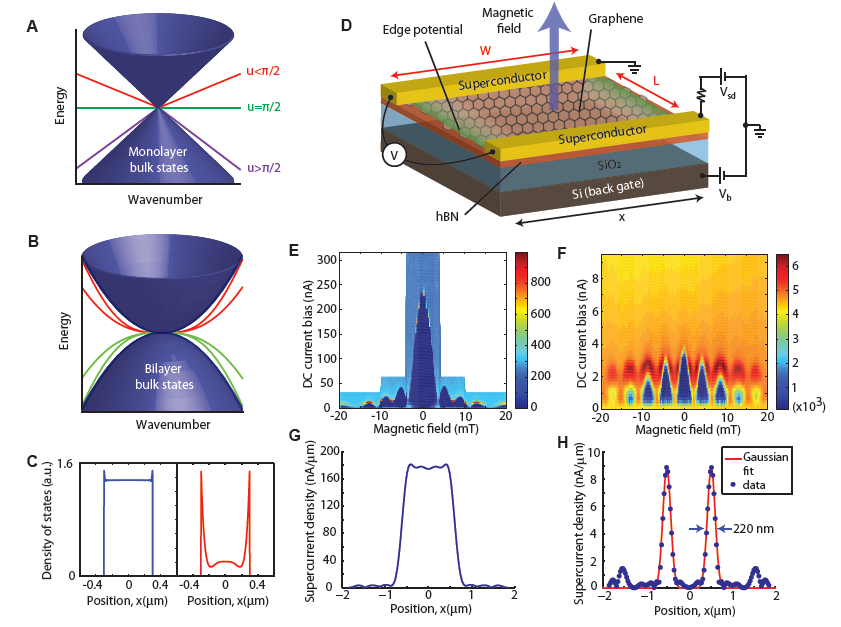}
\end{figure}

\newpage
\textbf{Figure 2}
\begin{figure}[!h]
\includegraphics[width=160mm]{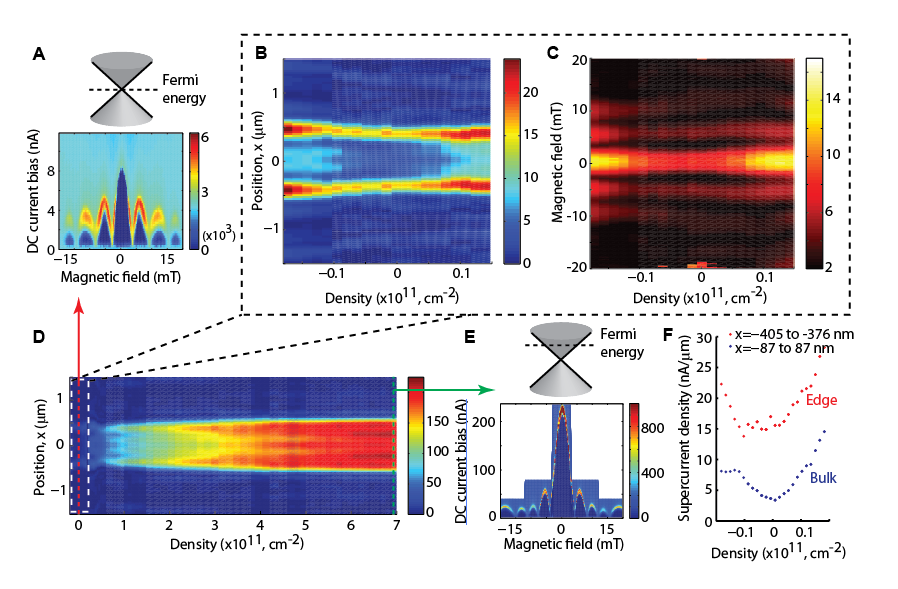}
\end{figure}

\newpage
\textbf{Figure 3}
\begin{figure}[!h]
\includegraphics[width=160mm]{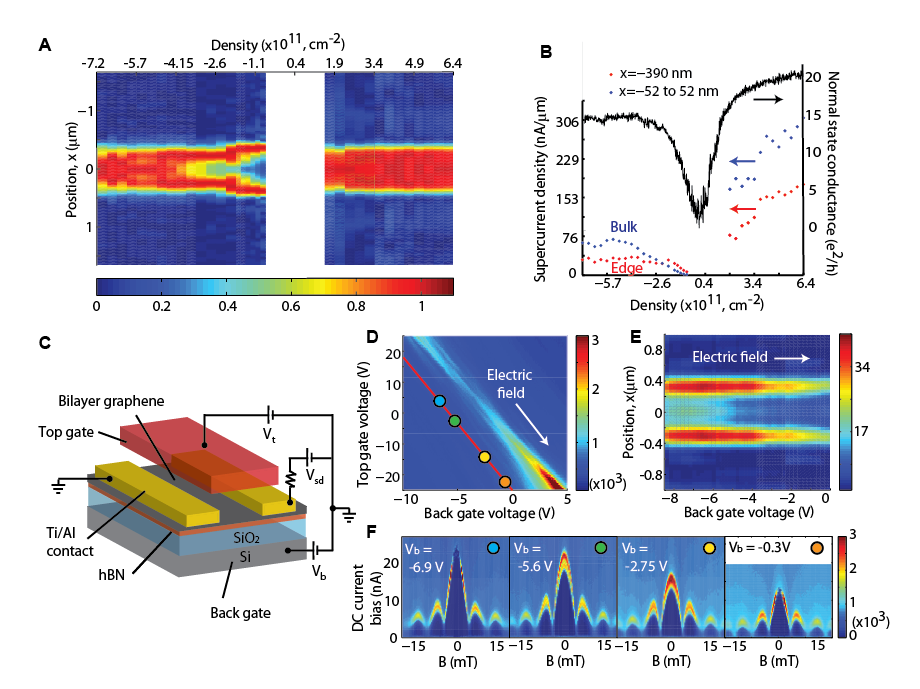}
\end{figure}

\newpage
\textbf{Figure 4}
\begin{figure}[!h]
\includegraphics[width=120mm]{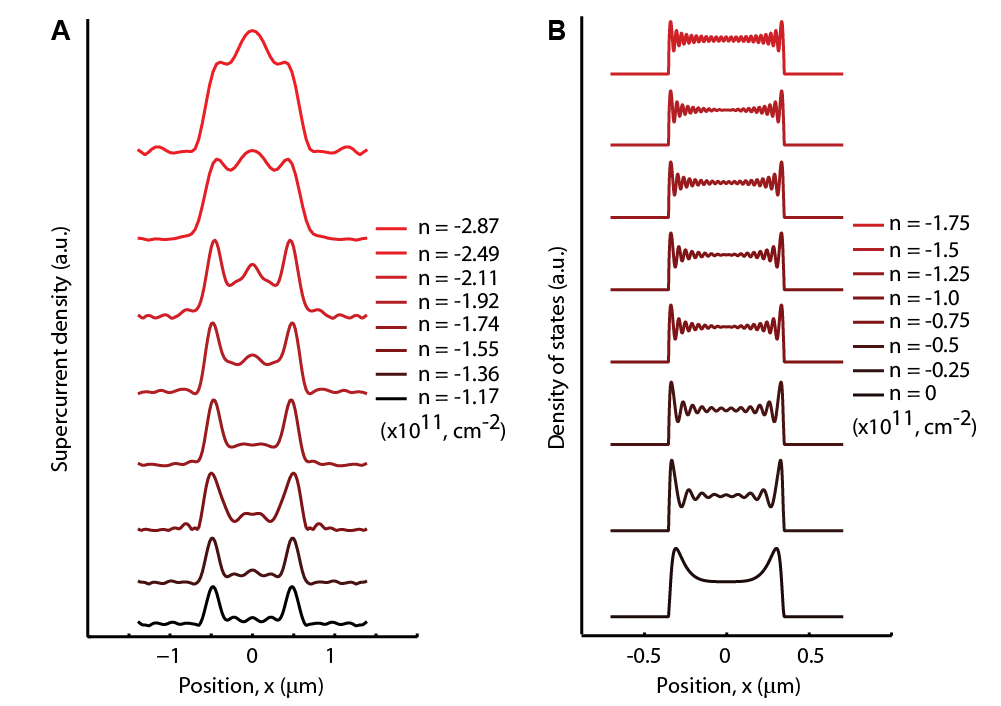}
\end{figure}


\begin{thebibliography}{10}

\bibitem{RMP1}
A.~H. Castro~Neto, F.~Guinea, N.~M.~R. Peres, K.~S. Novoselov, A.~K. Geim, {\it
  Rev. Mod. Phys.\/} {\bf 81}, 109 (2009).

\bibitem{Klein1}
M.~I. Katsnelson, K.~S. Novoselov, A.~K. Geim, {\it Nat. Phys.\/} {\bf 2},
  620– (2006).

\bibitem{Klein5}
V.~V. Cheianov, V.~Fal'ko, B.~L. Altshuler, {\it Science\/} {\bf 315}, 1252
  (2007).

\bibitem{Shytov1}
A.~V. Shytov, M.~S. Rudner, L.~S. Levitov, {\it Phys. Rev. Lett.\/} {\bf 101},
  156804 (2008).

\bibitem{Klein2}
A.~F. Young, P.~Kim, {\it Nat. Phys.\/} {\bf 5}, 222– (2009).

\bibitem{Klein4}
L.~Campos, {\it et~al.\/}, {\it Nat. Commun.\/} {\bf 3}, 1239 (2012).

\bibitem{Klein3}
A.~Varlet, {\it et~al.\/}, {\it Phys. Rev. Lett.\/} {\bf 113}, 116601 (2014).

\bibitem{Mayorov1}
A.~S. Mayorov, {\it et~al.\/}, {\it Nano Lett.\/} {\bf 11}, 2396 (2011).

\bibitem{MagneticFocusing1}
T.~Taychatanapat, K.~Watanabe, T.~Taniguchi, P.~Jarillo-Herrero, {\it Nat.
  Phys.\/} {\bf 9}, 225 (2013).

\bibitem{GuidedModes1}
J.~M. Pereira, V.~Mlinar, F.~M. Peeters, P.~Vasilopoulos, {\it Phys. Rev. B\/}
  {\bf 74}, 045424 (2006).

\bibitem{GuidedModes2}
F.-M. Zhang, Y.~He, X.~Chen, {\it Applied Physics Letters\/} {\bf 94} (2009).

\bibitem{GuidedModes3}
R.~R. Hartmann, N.~J. Robinson, M.~E. Portnoi, {\it Phys. Rev. B\/} {\bf 81},
  245431 (2010).

\bibitem{Klein7}
J.~R. Williams, T.~Low, M.~S. Lundstrom, C.~M. Marcus, {\it Nat. Nanotech.\/}
  {\bf 6}, 222 (2011).

\bibitem{Ritter1}
K.~A. Ritter, J.~W. Lyding, {\it Nat. Mater.\/} {\bf 8}, 235– (2009).

\bibitem{Crommie1}
C.~Tao, {\it et~al.\/}, {\it Nat. Phys.\/} {\bf 7}, 616 (2011).

\bibitem{Morpurgo2}
J.~B. Oostinga, H.~B. Heersche, X.~Liu, A.~F. Morpurgo, L.~M.~K. Vandersypen,
  {\it Nat. Mater.\/} {\bf 7}, 151 (2007).

\bibitem{Dominik1}
D.~Bischoff, F.~Libisch, J.~Burgdörfer, T.~Ihn, K.~Ensslin, {\it Phys. Rev.
  B\/} {\bf 90}, 115405 (2014).

\bibitem{Dean1}
C.~R. Dean, {\it et~al.\/}, {\it Nat. Nanotech.\/} {\bf 5}, 722 (2010).

\bibitem{TinkhamBook}
M.~Tinkham, {\it Introduction to Superconductivity\/} (McGraw-Hill Book Co.,
  New York, NY, 1975).

\bibitem{Dynes1}
R.~C. Dynes, T.~A. Fulton, {\it Phys. Rev. B\/} {\bf 3}, 3015 (1971).

\bibitem{DasSarma1}
H.~Y. Hui, A.~M. Lobos, J.~D. Sau, S.~D. Sarma  (2014).

\bibitem{SQUID1}
R.~C. Jaklevic, J.~Lambe, J.~E. Mercereau, {\it Phys. Rev. Lett.\/} {\bf 12},
  159 (1964).

\bibitem{Blake1}
P.~Blake, {\it et~al.\/}, {\it Solid State Comm.\/} {\bf 149}, 1068 (2009).

\bibitem{Kern1}
E.~J.~H. Lee, K.~Balasubramanian, R.~T. Weitz, M.~Burghard, K.~Kern, {\it Nat.
  Nanotech.\/} {\bf 3}, 486– (2008).

\bibitem{Wu1}
Y.~Wu, {\it et~al.\/}, {\it Nano Lett.\/} {\bf 12}, 1417 (2012).

\bibitem{Gap1}
E.~McCann, {\it Phys. Rev. B\/} {\bf 74}, 161403 (2006).

\bibitem{Gap2}
E.~V. Castro, {\it et~al.\/}, {\it Phys. Rev. Lett.\/} {\bf 99}, 216802 (2007).

\bibitem{Jackiw1}
R.~Jackiw, C.~Rebbi, {\it Phys. Rev. D\/} {\bf 13}, 3398 (1976).

\bibitem{Bandbending1}
P.~G. Silvestrov, K.~B. Efetov, {\it Phys. Rev. B\/} {\bf 77}, 155436 (2008).

\bibitem{Zigzag1}
K.~Nakada, M.~Fujita, G.~Dresselhaus, M.~S. Dresselhaus, {\it Phys. Rev. B\/}
  {\bf 54}, 17954 (1996).

\bibitem{Akhmerov1}
A.~R. Akhmerov, C.~W.~J. Beenakker, {\it Phys. Rev. B\/} {\bf 77}, 085423
  (2008).

\bibitem{Guinea1}
E.~V. Castro, N.~M.~R. Peres, J.~M. B.~L. dos Santos, A.~H.~C. Neto, F.~Guinea,
  {\it Phys. Rev. Lett.\/} {\bf 100}, 026802 (2008).

\bibitem{Muller1}
P.~Y. Huang, {\it et~al.\/}, {\it Nature\/} {\bf 469}, 389 (2011).

\bibitem{Mele1}
F.~Zhang, A.~H. MacDonald, E.~J. Mele, {\it PNAS\/} {\bf 110}, 10546 (2013).

\bibitem{Alden1}
J.~S. Alden, {\it et~al.\/}, {\it PNAS\/} {\bf 110}, 11256 (2013).

\end{thebibliography}
\end{document}


\begin{center}
\huge{{Supplementary Materials for}}\\[0.5 cm]
\Large{\textbf{Spatially resolved edge currents and guided-wave electronic states in graphene}}\\[0.5 cm]
\large{M. T. Allen, O. Shtanko, I. C. Fulga, A. Akhmerov, K. Watanabi, T. Taniguchi,  P. Jarillo-Herrero, L. S. Levitov, and A. Yacoby$^{*}$} \\[0.5 cm]

$^{*}$Correspondence to: yacoby@physics.harvard.edu\\[1.0 cm] 
\end{center}



\large{\textbf{This PDF file includes:}}\\

Materials and Methods

Figs. S1 to S5

Table S1

References (36-37)

\newpage
\textbf{Materials and Methods}\\

\underline{Modeling electronic guided modes}\\

A full model of supercurrent-carrying states in our system should account, in principle, for a number of microscopic effects.
This includes, in particular, the microscopic details of transport through the NS interfaces, the realistic edge potential profile due to band bending near graphene edge, as well as  the effects of disorder. Since treating all these issues simultaneously and on equal footing makes such a modeling a daunting task, here we resort to some simplifications. First, we will completely ignore the effects of induced superconductivity, 
focusing on the normal metallic state of a pure graphene. Second, we consider a clean system and account for disorder scattering perturbatively at the end. Third, since states in a clean system, being delocalized, are capable of carrying supercurrent, we will focus on evaluating the density of states (DOS) taking it to reflect on the current-carrying capacity of the system. Of course, such an approach may be questioned for disordered systems in which some states are localized, and therefore can contribute to DOS but not to supercurrent. However, taking into account that in a clean system all states possess a roughly similar current-carrying capacity, we adopt this approximation on the merit of its simplicity.\\

Turning to the discussion of system geometry, we note two points. First, as discussed in the main text, the problem of guided states on a halfplane near the edge $x>x_0$ can be mapped onto a similar problem on a full plane by accounting for the states in valleys $K$ and $K'$ mixing at the edge. This mapping is particularly transparent for the armchair edge, where the boundary condition for the spinor wavefunctions in the two valleys is simply $\psi_{K}+\psi_{K'}=0$. In this case, one can see that the two-valley half-plane problem is mathematically equivalent to the problem posed on a full plane for particles in just one valley, provided the line potential for the latter problem is taken to be a sum of the original edge potential and its mirror-reflected double, $V(x>x_0)\,\to\,V(|x-x_0|)$. \\

Second, the states with the wavelengths larger than the edge potential width can be described by a delta function approximation. In that, a realistic microscopic potential $V(x)$ is replaced by a delta-function pseudopotential $\tilde V(x)=\lambda\delta(x-x_0)$, where $\lambda=\int V(x')dx'$ and $x_0$ is the edge position. For a system of width $w$ with two parallel edges positioned at $x_0=\pm w/2$ we therefore arrive at the model  
\be \label{model}
V(x) = \lambda \delta(x+w/2)+\lambda \delta(x-w/2), 
\ee
with $-\infty<x<\infty$. Carriers in this system are described by the massless Dirac Hamiltonian
\be \label{hamiltonian}
H =H_0+V(x),\quad H_0=v\sigma_1 p_x+v\sigma_2 p_y
\ee
with $v\approx 10^6{\rm m/s}$ the carrier velocity and $\sigma_{1,2}$ the pseudospin Pauli matrices. 

As stated above, we will use spatially-resolved DOS for the problem (\ref{hamiltonian}) as a measure of current-carrying capacity of the system. In justification we note that an electron system carrying normal electric current can be understood in terms of changes in the occupancy of the states near the Fermi energy. As a result, the spatially-resolved current density will vary in the same manner as DOS
%
%
\be
N(\mu,\vec r) = \frac{dn(\vec r)}{d\mu}
,\quad
n(\vec r)=\la \psi^\dagger(\vec r)\psi(\vec r)\ra
.
\ee
Here $n$ is the total carrier density and $\mu$ is chemical potential.
Below we evaluate DOS as a function of position and energy, focusing on the characteristic features due to the guided modes. \\

\begin{figure*}[h!]
\begin{minipage}[h!]{0.49\linewidth}
\center{\includegraphics[width=0.9\linewidth]{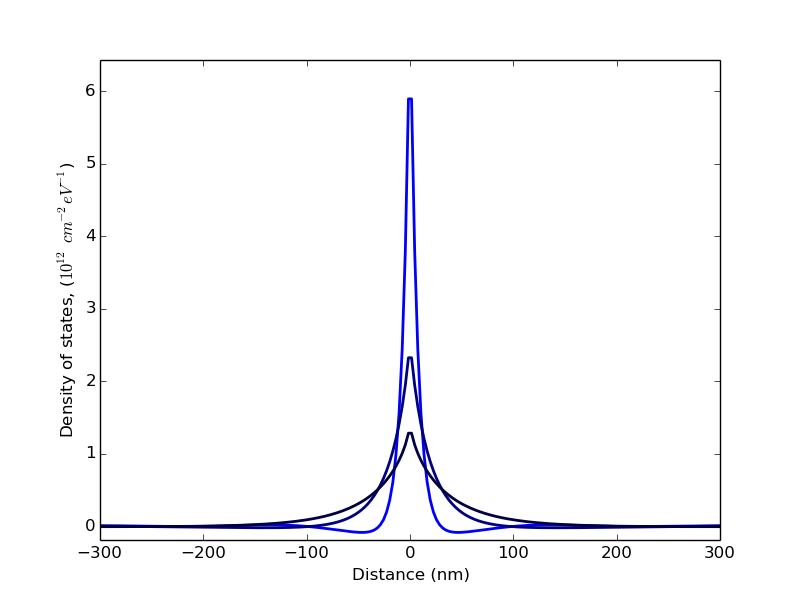}}
\end{minipage}
\hfill
\begin{minipage}[h!]{0.49\linewidth}
\center{\includegraphics[width=0.9\linewidth]{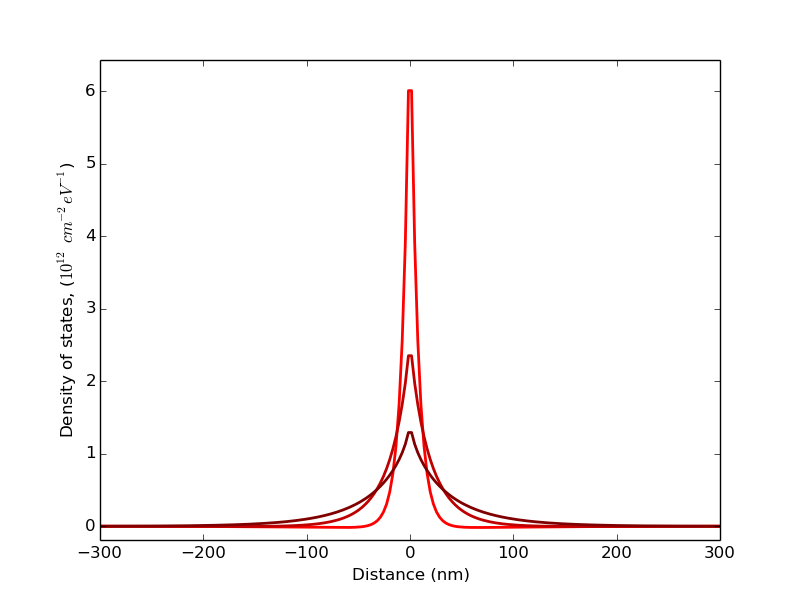}}
\end{minipage}

\caption{}
\label{peaks}
\end{figure*}

Taking into account that typical wavelength values of relevant electronic states, $\lambda \sim 10$nm , are much smaller than the distance between edges $w\sim \,1{\rm \mu m}$, we can represent DOS
in the form
\be
\label{full_density}
N(\mu,x) = N_0(\mu)+N_1(\mu,x-w/2)+N_1(\mu,x+w/2)
\ee
where $N_0$ is the DOS of a uniform infinite system, 
\be
\label{n_bulk}
N_0(\ve) = \frac{|\ve|}{2\pi \hbar^2v^2}
\ee
and $N_1$ is the contribution to DOS from a single delta-function line potential, placed at $x=0$. Below we derive an expression
\be
\label{n_edge}
N_1(\ve,x) = \frac{4\lambda}{\pi \hbar v}\,\Im\int\frac{dp}{2\pi}\frac{p^2 e^{-2\kappa_{\ve,p}|x|/\hbar}}{\kappa_{\ve,p}
\bigl[ 4\lambda\ve+(4-\lambda^2)\hbar v\kappa_{\ve,p}\bigl]}
,\quad 
\kappa_{\ve,p} = \sqrt{p^2-(\epsilon/\hbar v)^2}
\ee
where the energy $\ve$ is taken to have an infinitesimal positive imaginary part. In the final result for DOS $\ve$ must be replaced by the chemical potential, $\ve = \mu$. 
The spatial dependence described by Eq.\eqref{n_edge} is shown on Fig. \ref{peaks}.\\

In our model, which is essentially non-interacting, the effects of screening can be included {\it ad hoc} by treating the potential strength in Eq.\eqref{model} as a function of carrier density. 
Since the latter is parameterized by $\mu$, we will use a simple model 
\be \label{eq:screening}
\lambda\rightarrow\lambda'=\frac{\lambda}{1+(|\mu|/\mu_0)^\alpha}
\ee
where the parameter $\mu_0$ depends on microscopic details. Comparing to the data indicates that a reasonably good fit can be achieved for $\alpha \approx 2$.\\

Modeling results are presented in \addOS{Fig.1(c) of the main text} for energies corresponding to carrier densities $n = 0.05\cdot 10^{11} \text{cm}^{-2}$ (red curve) and $n = 2.5\cdot10^{11} \text{cm}^{-2}$ (blue
curve), where we evaluated $n$ accounting for the spin and valley degeneracy in graphene. Potential strength is chosen to be
$\lambda = - 1.5\,\hbar v \approx 1$ eV$\cdot$nm and the screening parameter value is $\mu_0 = 0.2\sqrt{\pi\hbar^2v^2n_0} \approx 7\,\text{meV}$, where $n_0 = 10^{11} \text{cm}^{-2}$ is the corresponding scale for density. \\


The simulation for graphene bilayer, which was used to generate \addOS{Fig.1(b) and Fig. 4(b) of the main text}, was carried out using an effective delta function potential approximation, as above. Greens function expressed through the T-matrix was used to obtain mode dispersion and DOS in a manner similar to our treatment of modes in a single layer. For the delta function strength we used the best-fit value $\lambda = 0.5 $ eV$\cdot$nm (and no screening).


\underline{Microscopic derivation}\\

Here we consider long-wavelength modes for a potential line positioned at $x=0$. This problem is described by the Hamiltonian \eqref{hamiltonian} with  $V(x) = \lambda \delta(x)$. For this problem we construct the Greens function which takes the full account of scattering by the potential. 
As is well known, the discrete spectrum of the system (in our case, the guided modes) can be conveniently expressed through the poles of the electron Greens function. Likewise, the spatially-resolved DOS is expressed as the Greens function trace. The Greens function, in turn, can be straightforwardly evaluated using Dysons's equation and the T-matrix representation:
\be\label{eq:G}
G = G_0+G_0 V G_0+G_0  V G_0 V G_0 + \dots 
=G_0+G_0 T G_0
\ee
where $G_0= (i\epsilon-H_0)^{-1}$. 

Assuming that the 
phase and amplitude of the electron wavefunction are given by a continuous function of $x$, we can express the quantity $T$ as
\be\label{eq:T_general}
T(\epsilon, p_y) =\lambda\Bigl(1-\lambda\int\frac{dp_x}{2\pi\hbar}G_0(\vec p)\Bigl)^{-1}
\ee
The continuity assumption should in practice be relaxed by a weaker assumption accounting for the phase jump of the wavefunction across the delta function potential at $x=0$ (see main text). Here, however, we will proceed with Eq.\eqref{eq:T_general} on the merit of its simplicity. 
Evaluating the integral in Eq.\eqref{eq:T_general} gives
\be\label{eq:T} 
T(\epsilon, p_y) = \lambda\Bigl(1+\frac{\lambda }{2\hbar v}\lp i\tilde\epsilon+\sigma_1\tilde p\rp\Bigl)^{-1}
\ee
where we defined 
\be 
\tilde\epsilon=\frac{\epsilon}{\sqrt{\epsilon^2+v^2p_y^2}}\qquad \tilde p=\frac{\hbar v p_y}{\sqrt{\epsilon^2+\hbar^2v^2p_y^2}}
\ee
Here $\ve$ is the Matsubara frequency, with a suitable analytic continuation $i\ve\to\ve+i0$ to be performed at the end.\\

The T-matrix poles  give the guided modes dispersion
\be\label{eq:dispersion}
\epsilon=\pm \hbar u|p_y|,\quad u=v\frac{4 \hbar^2v^2-\lambda^2}{4 \hbar^2v^2+\lambda^2}
\ee
where the sign is given by $\pm=\sgn\lambda$. 
Since $|u|<v$, the energies $\epsilon=\pm u |p_y|$ are positioned,  for each $p_y$ value,  outside the Dirac continuum of the bulk states.
This expression behaves in a qualitatively similar way to the exact dispersion  derived in the main text, Eq.(1) \addOS{[see Fig.1(a) of the main text]}. The guided modes described by Eq.\eqref{eq:dispersion} are quasi-1D states that propagate as plane waves in the $y$ direction along the $x=0$ line and
decay exponentially as evanescent waves in the transverse direction.\\

Spatially-resolved DOS can be evaluated as
\be\label{eq:n(E)}
N(\epsilon,\vec r)=-\frac1{\pi}{\rm Im}\,\Tr G(\epsilon, \vec r,\vec r')_{\vec r=\vec r'}
\ee
where the energy variable is analytically continued from positive imaginary to real values via $i\epsilon\to\epsilon+i0$ and a trace is taken over pseudospin variables. 
To proceed with our calculation, we will need Greens function evaluated in a mixed position-momentum representation 
\begin{align}
&G_0(\epsilon,p_y,x)=\int \frac{dp_x}{2\pi}e^{ip_x x}G_0(\epsilon,\vec p)
\\\nonumber
&= \frac{-i\tilde\epsilon-\sigma_2 \tilde p-i\sigma_1\sgn(x)
}{2\hbar v} \exp\bigl(-\kappa(i\ve) |x|/\hbar\bigl)
\end{align}
where $\kappa(i\ve)=\sqrt{(\epsilon/\hbar v)^2+p_y^2}$. 

The  trace of an equal-point Greens function in Eq.(\ref{eq:n(E)}) then could be evaluated from Eq.\eqref{eq:G} with the help of Eq.\eqref{eq:T}:
\be\label{eq:TrG} 
\Tr G(\epsilon,x'=x)=\sum_{p_y}\Biggl(  
\frac{\tilde\epsilon}{i\hbar v}
+\frac{4\lambda \tilde p^2 e^{-2\kappa |x|/\hbar}}{\hbar v\lb\lp 2+i\lambda\tilde\epsilon\rp^2-\lambda^2\tilde p^2\rb}\Biggl)
\ee
where the two terms represent contributions of $G_0$ and $G_0VG_0$, respectively. \\

As a warmup, we consider the first term of \eqref{eq:TrG}. 
Introducing a UV cutoff $p_0 = \ve_0/\hbar v$ we evaluate the sum over $p_y$ as
\be
 \int_{-p_0}^{p_0}\frac{dp_y}{2\pi\hbar}\frac{\epsilon}{\sqrt{\epsilon^2+\hbar^2v^2p_y^2}}=\frac{\epsilon}{\pi \hbar v}\ln\frac{\epsilon_0}{\epsilon}
.
\ee
Performing analytic continuation $\epsilon\to \delta-i\epsilon$, we arrive at
\be
N_0(\ve) = - \frac{\epsilon}{\pi^2 \hbar^2v^2}\Im\ln\frac{\epsilon_0}{\delta-i\epsilon} 
\ee
where $\delta=+0$. Taking the imaginary part, we obtain the expression in Eq.\eqref{n_bulk}.\\

Next, we proceed to evaluate the second term in Eq.\eqref{eq:TrG}. Performing the same analytic continuation, we arrive at the result in Eq.\eqref{n_edge}. The expression in Eq.\eqref{n_edge} can be conveniently analyzed by dividing the integral  into two parts, taken over the domains $|p_y|>\ve/\hbar v$ and $|p_y|<\ve/\hbar v$, respectively. The latter contribution is particularly simple because it is governed by the pole \eqref{eq:dispersion} and can be easily evaluated, giving
\be
\label{guided}
N_{\rm g.w.}(\ve,x)=\frac{2\epsilon\lambda  }{\hbar^2vu(4-\lambda^2)} 
e^{-2 \sqrt{(v/u)^2-1} |x| |\epsilon|/\hbar v}
\ee
This contribution is solely due to the guided edge mode.  As illustrated in the Fig \ref{peaks}, 
this term dominates the peak structure in DOS for guided waves. \\

We used the full expression in  Eq.\eqref{n_edge} to produce the spatially-resolved DOS curves shown in \addOS{Fig.1(c) of the main text}. In that, we accounted for screening, as described in Eq.\eqref{eq:screening}. Because of screening, the peak structure is more prominent at low chemical potential, and is suppressed relatively to the bulk DOS at high chemical
potential values.\\

\underline{The effect of disorder}\\


Here we estimate the disorder scattering rate $\gamma(k)$ for guided modes [see Eq.(1) in the main text and accompanying discussion]. We will model edge roughness by a fluctuating delta function strength, treating the fluctuations as a gaussian white noise:
\be\label{eq:V+dV}
V(x,y)=(\lambda +\delta\lambda(y))\,\delta(x)
,\quad
\la \delta\lambda(y)\delta\lambda(y')\ra =\alpha\delta(y-y')
.
\ee 
Writing the Greens function as a series in the potential $V+\delta V$, Eq.(\ref{eq:V+dV}), we have 
\be
G=G_0+G_0(V+\delta V)G_0+G_0(V+\delta V)G_0(V+\delta V)G_0+...
\ee
Averaging the Greens function over disorder, we only need to account for the pair correlators $\la \delta\lambda(y)\delta\lambda(y')\ra$. 
In a non-crossing approximation, we express the disorder-averaged Greens function through a suitable self-energy
\be
\la G\ra = G_0+ G_0(V+\Sigma)G_0+ G_0(V+\Sigma)G_0(V+\Sigma)G_0+... 
\ee
where 
\be\label{eq:Sigma}
\Sigma(\epsilon)=\alpha
 \int \frac{dp_x}{2\pi} G(\epsilon,p_y,x,x')_{x=x'=0}
\ee
The quantity (\ref{eq:Sigma}) is complex-valued, with the imaginary part expressed through the density of states at $x=0$ as 
\be
{\rm Im}\,\Sigma(\epsilon)=-\pi \alpha N(\epsilon)_{x=0} 
\ee
The disorder scattering rate for the guided waves can now be found from the dispersion relation obtained from the T-matrix pole, Eg(\ref{eq:T_general}), 
which is corrected by the presence of $\Sigma$ as follows
\be \label{eq:dispersion_scatter}
1+(\lambda+\Sigma(i\epsilon))\frac{i\tilde\epsilon+\sigma_1\tilde p}{2\hbar v}=0
.
\ee
Here we continue to use Matsubara notation, as in Eqs.(\ref{eq:T_general}),(\ref{eq:T}).

Since the density of states scales linearly with energy, $N(\epsilon)\sim |\epsilon|$, we can solve Eq.\eqref{eq:dispersion_scatter} in the long-wavelength limit treating $\Sigma(i\epsilon)$ as a perturbation.  Writing $\epsilon=\epsilon_0(p_y)+\delta\epsilon$, where $\epsilon_0=u|p_y|$ is a solution for $\Sigma=0$, we linearize in $\delta\epsilon$ to obtain
%
\be
\delta\ve = -\frac1\lambda\Bigl(1-\frac{u^2}{v^2}\Bigl)\Sigma(i\epsilon_0)|p_y|
\ee
After analytic continuation, we obtain
\be
\gamma(p_y)  = \frac{\pi\alpha}{|\lambda|}\Bigl(1-\frac{u^2}{v^2}\Bigl)|p_y|N\bigl(u|p_y|\bigl)_{x=0}
\ee
Accounting for the linear scaling $N\sim|\epsilon|$, we find that the damping rate scales as a square of $p_y$, 
\be
\gamma(p_y) =\frac{\lambda}{\hbar^2 v (4-\lambda^2)}p_y^2
\ee
at small $p_y$. A similar dependence, albeit with a different prefactor, is found at large $p_y$. 

From this we conclude that the modes are undamped over lengthscales $\sim\lambda^2/\xi$, where $\lambda$ is a wavelength and $\xi$ is a disorder lengthscale. Taking realistic values $\lambda\approx 10-100\,{\rm nm}$ and $\xi\approx 0.1\,{\rm nm}$, we obtain an estimate for the guided mode mean free path in the $1-10\,{\rm \mu m}$ range. These large values can be traced to the weak confinement of the waves at small $p_y$. The weak confinement results in the mode wavefunction positioned mostly outside the confining potential, which reduces the impact of scattering. The mean free path rapidly grows with wavelength, in a direct analogy with guided optical waves in weakly guiding fiber designs, where weak confinement is employed to achieve exceptionally long mean free paths. \\

\underline{Josephson junctions: Device overview}\\

We analyze five graphene Josephson junctions on hBN with widths ranging from $W=800-1200$ nm and lengths ranging from $L=250-350$ nm (see Fig. 1d for a labeled device schematic).  Listed in Table S1 are details on individual sample geometries.  The small $L/W$ aspect ratios place these devices are in the narrow junction limit, where the the critical current $I_c$ can be approximated as a phase dependent summation over many parallel 1D current channels (Equation 2 in the main text).   Electrical measurements are conducted using standard Lockin techniques in a Leiden Cryogenics Model Minikelvin 126-TOF dilution refrigerator with a base temperature of 10 mK, well below the critical temperature of Al. \\


Using a dry transfer method, graphene/hBN stacks are sequentially deposited on a 300 nm thermally grown SiO$_2$ layer, which covers a doped silicon substrate functioning as a global back gate.  Graphene flakes are etched to the desired geometry using a 950 PMMA A4 polymer mask ($\sim 200$ nm thick; spun at 4000 rpm) followed by an RIE O2 plasma etch. Titanium/aluminum (Ti/Al) superconducting electrodes are defined on selected flakes using electron beam (ebeam) lithography on a 950 PMMA A4 resist mask, followed by thermal evaporation and liftoff in acetone.  For the titanium adhesion layer, we evaporate 10 nm at a rate of 0.3 Angstrom/s.  This is followed by an evaporation of a 70 nm aluminum layer at a rate of 0.5 Angstrom/s at pressures in the low to mid $10^{-7}$ Torr range. For dual-gated bilayers, suspended top gates are fabricated using a standard PMMA/MMA/PMMA trilayer resist method which leaves a 200 nm air gap between the top gate and graphene.  After using ebeam lithography to define the gates, which employs position-dependent dosage, Cr/Au (3/425 nm) gates are deposited using thermal evaporation and liftoff in acetone.  To remove processing residues and enhance quality, devices were current annealed in vacuum at dilution refrigerator temperatures.  We note that edge currents were detected both in current-annealed and intrinsically high quality non-annealed devices; typically the appearance of edge currents coincided with the occurrence of Fabry-Perot interference in the ballistic transport regime.  All five graphene Josephson junctions exhibit similar transport behavior.  Additional data sets are provided in the Supplementary Figures.\\


\underline{Fourier method for extraction of supercurrent density distribution}\\

In a magnetic field $B$, the critical current $I_c(B)$ through a Josephson  junction equals the magnitude of the complex Fourier transform of the current density distribution $J(x)$:
\begin{equation}
I_c(B)=|\mathcal{I}_c(B)|=\left| \int_{-\infty}^{\infty} J(x) \exp(2\pi i(L+l_{Al})Bx/\Phi_0) dx \right|
\end{equation}
where $x$ is the dimension along the width of the superconducting contacts (labeled in Fig. 1d), $L$ is the distance between contacts, $l_{Al}$ is the magnetic penetration length (due to a finite London penetration depth in the superconductor and flux focusing), and $\Phi_0=h/2e$ is the flux quantum.  Relevant in the narrow junction limit where current is only a function of one coordinate, Equation (28) provides a simple and concise description of our system.  We employ Fourier techniques introduced by Dynes and Fulton to extract the real space current density distribution from the magnetic interference pattern $I_c(B)$.  By expressing the current density as $J(x)=J_{s}(x)+J_{a}(x)$, where $J_{s}(x)$ and $J_{a}(x)$ are the symmetric and antisymmetric components, the complex critical current can be rewritten as:
\begin{equation}
\mathcal{I}_c(B)= \int_{-\infty}^{\infty} 
J_s(x)\cos(2\pi (L+l_{Al})Bx/\Phi_0) dx
+ i\int_{-\infty}^{\infty} 
J_a(x)\sin(2\pi (L+l_{Al})Bx/\Phi_0) dx
\end{equation}
We calculate symmetric component of distribution, the relevant quantity for analyzing edge versus bulk behavior, as the antisymmetric component goes to zero in the middle of the sample.  For symmetric solutions, $\mathcal{I}_c(B)$ is purely real.  To reconstruct $\mathcal{I}_c(B)$ from the measured critical current, the sign of $I_c(B)$ is reversed for alternating lobes of the Fraunhofer interference patterns.  The extracted supercurrent distribution is expressed as an inverse Fourier transform:
\begin{equation}
J_s(x)\approx\int_{-\infty}^{\infty} \mathcal{I}_c(B)
\exp(2\pi i(L+l_{Al})Bx/\Phi_0) dB
\end{equation}
Because $I_c(B)$ is only nonzero over a rectangular window dictated by the finite scan range $B_{min}<B<B_{max}$, distribution extracted numerically is given by the convolution of $J(x)$ with the sinc function.  To reduce artifacts due the convolution, we employ a raised cosine filter to taper the window at the endpoints of the scan.  Explicitly,
\begin{equation}
J_s(x)\approx\int_{B_{min}}^{B_{max}} \mathcal{I}_c(B) 
\cos^n(\pi B /2 L_B)
\exp(2\pi i(L+l_{Al})Bx/\Phi_0) dB
\end{equation}
where $n=0.5-1$ and $L_B=(B_{max}-B_{min})/2$ is the magnetic field range of the scan.\\

\underline{Gaussian fits to extract edge state widths}\\

To extract a length scale for the width of the edge currents near the Dirac point, we fit the experimental supercurrent density distribution $J_c(x)$ to the Gaussian function
\begin{equation}
J_c^G(x)= b\left( \exp\left( \frac{-(x-a)^2}{c} \right) +\exp\left( \frac{-(x+a)^2}{c} \right) \right)
\end{equation}
where $a$ determines the spatial peak offset, $b$ determines peak height, and $c$ determines peak width.  For the data in Fig. 1H, the fit parameters are $a=0.515$, $b=8.8$, and $c=0.017$.  The effective edge current width, given by the Gaussian full width at half maximum $x_{FWHM}=2\sqrt{c\cdot\ln2}$, is 220 nm.\\

\underline{Edge versus bulk amplitudes}\\

To more quantitatively assess the evolution of edge and bulk currents with electronic carrier density $n$, we plot line cuts of the individual contributions (see Fig. 2f and 3b). These are given by:
\begin{equation}
J_c^{edge}(n)=\sum_{x_i=-x_W}^{-x_W+\epsilon_1}\frac{J_c(x_i,n)}{N_1} \quad \mathrm{and} \quad
J_c^{bulk}(n)=\sum_{x_i=-\epsilon_2}^{\epsilon_2}\frac{J_c(x_i,n)}{N_2}
\end{equation}
for a graphene flake whose full width spans from $-x_W$ to $x_W$.  $J_c^{edge}(n)$ is the spatially averaged current amplitude over a small window of width $\epsilon_1$ from the edge of the flake.  Similarly, $J_c^{bulk}(n)$ is the spatially averaged current amplitude over a strip of width $2\epsilon_2$ around the center of the flake.  $N_1=\epsilon_1/x_{step}$ and $N_2=\epsilon_2/x_{step}$, where $x_{step}$ is the distance between data points (determined by the magnetic field range of the scan).  For example, for the plots in Fig. 2F, $x_W=405$ nm, $\epsilon_1= 29$ nm, and $\epsilon_2=87$ nm.\\

Based on the edge versus bulk current profiles, one may infer whether edge doping is the dominant cause of edge currents in our devices.  In the presence of edge doping, the edge versus bulk contributions should be reversed for opposite polarities of bulk carriers (for example, edge dominated behavior at high densities on the electron side and bulk dominated behavior at high densities on the hole side), which is not consistent with the data.  Bulk-dominated or flat distributions appear at both high electron and hole doping fairly consistently.  As a second test, one can track the edge versus bulk contributions through the Dirac point to detect an offset in gate voltage between the charge neutrality point at the edge versus in the bulk.  We did not detect positional density offset substantial enough to account for the large edge currents in these devices (Fig. 2F).\\

\underline{Bayesian method for extraction of supercurrent density distribution}\\

The critical current as a function of the magnetic field, $I_c(B)$, is related to the current density through the junction, $J_c(x)$, as

\begin{equation}\label{eq:ij}
 I_c(B) = \int_{-\frac{W}{2}}^{\frac{W}{2}} dx\,J_c(x) \exp \left( 2\pi ix LB/\Phi_0 \right),
\end{equation}
with $L$ and $W$ the length and width of the junction, and $\Phi_0=h/2e$ the superconducting flux quantum.\\

In the measured $|I_c(B)|$ all information about its complex phase is lost, making the problem of determining the current density not have a unique solution. Using the method of Dynes and Fulton (DF), a unique solution can be found under the assumption of a symmetric current distribution, $J_c(x)=J_c(-x)$. In practice however, disorder and inhomogeneities in the junction will lead to asymmetric current densities. Additionally, since experiments are performed over a finite range of magnetic fields, there is a cutoff in the current density resolution. Neither this finite resolution, nor experimental uncertainties are taken into account in the DF method, meaning it can only provide a qualitative estimate of $J_c(x)$.\\

To gain a more quantitative understanding, we instead ask what is the distribution of $J_c(x)$ which produces the same critical current $I_c(B)$. We answer this question by performing Bayesian inference to obtain the posterior distribution of the current density, given the measured critical current. In our case, Bayes' rule reads:

\begin{equation}\label{eq:bayes}
  P ( J_c ; |I_c| ) = \frac{ P( |I_c| ;J_c ) P (J_c )}{ P (|I_c| )}.
\end{equation}

Here, $ P ( J_c ; |I_c| )$ is the posterior distribution of the current density, the quantity we want to calculate, while $ P (J_c )$ is its prior distribution. The likelihood function $ P( |I_c| ;J_c )$ indicates the compatibility of the measured critical current with a given current density:

\begin{equation}
 P( |I_c| ;J_c ) = \exp \left[ -\frac{(|I_c| - |I_c^f|)^2 }{2\varepsilon^2} \right],
\end{equation}
where $I_c^f$ is the current obtained from $J_c$ by using Eq.~\eqref{eq:ij}, $I_c$ is the measured current, and $\varepsilon$ is the measurement error. The factor $P (|I_c| )$ is the same for all current densities, meaning it does not enter into determining their relative probabilities.\\

The experimental current profiles are extracted from scans of the differential resistance as a function of DC current bias and magnetic field, $dV/dI(I_{\rm DC}, B)$. Within the same scan, for some field values $dV/dI$  has a clear maximum, while for others it monotonically increases towards its normal state value. We extract the critical current as the value $I_{\rm DC}$ at which the differential resistance is $x \times {\rm max}\,dV/dI$, choosing a value of $x\lesssim 1$. This selects points close to the maxima at field values where they are well defined, and points close to where the differential resistance reaches its normal state value otherwise. The uncertainty is obtained in the same fashion, by choosing a slightly smaller cutoff.\\

We maximize the likelihood function using a Monte Carlo sampling algorithm\cite{pymc}. 
To get a large resolution of the current density without a significant increase in the dimensionality of the sampling space, we expand $J_c(x)$ as

\begin{equation}\label{eq:jc}
 J_c(x) = \sum_{n=0}^{N} A_n \cos(2\pi n x/L)
\end{equation}
and enforce $J_c(x)>0$ for all $x$. The $A_n$ coefficients determine the shape of the distribution, which in Eq.~\eqref{eq:jc} is assumed to be symmetric, $J_c(x)=J_c(-x)$. Using an asymmetric form would typically lead to a critical current which shows node lifting -- the minima of $I_c(B)$ have nonzero values. While this feature is present in the measured critical current, it can be accounted for by factors other than an asymmetric current distribution \cite{Heida1998}, 
such as relatively small aspect ratios ($\sim$5), and a non-sinusoidal current-phase relationship arising from a large junction transparency.  Using a symmetric $J_c$ avoids this ambiguity, and has the additional advantage of providing a more direct comparison between our method and that of Dynes and Fulton.\\

The likelihood function is maximized by allowing the $A_n$ coefficients to vary at each Monte Carlo step. As $N$ is increased the posterior distribution of the current density widens, an indication of over-fitting. This increase in uncertainty serves as a criterion for choosing $N$, which for the typical dataset is between 4 and 8. The priors of $A_n$ are set to the uniform distribution $[-\max(I_c), \max(I_c)]$.\\

An example of our method is shown in Fig.~\ref{fig:jcic}, using $N=5$. The current density is peaked at the edges of the sample, a feature also recovered in the DF approach. The corresponding critical current is in good agreement with the measured one, with the exception of the regions close to the nodes. Fig.~\ref{fig:jcic} indicates that the supercurrent through the junction flows mainly along its edges. As a further test of the edge state contribution, we modify the functional form of the current density in Eq.~\eqref{eq:jc}, to explicitly allow for edge states. We add delta functions to the current density at the edges of the sample, $J_c(x) \rightarrow J_c(x) + d_L \delta(x+W/2) + d_R \delta (x-W/2)$, and estimate the contribution of edge states as the ratio of $d_L+d_R$ to the total current density $J_c^{\rm tot}$. As the carrier density approaches zero a significant fraction of the supercurrent is carried by the edge states, with $(d_L+d_R)/J_c^{\rm tot} \simeq 0.45$ (see Fig.~\ref{fig:deltas}).

\newpage
\textbf{Supplementary Figures}\\

\textbf{Fig. S1}. 

The excess contribution to the spatially-resolved DOS near a line delta function potential, $\Delta N(\epsilon,x)=N(\epsilon,x)-N_0(\epsilon)$ vs. distance from the delta function.  Subtracted is the bulk contribution $N_0$ given in Eq.(S5).  The left panel shows the full excess contribution obtained from Eq.(S6), the right panel shows the contribution solely due to the guided modes, Eq.(S18).  The two contributions are nearly identical, confirming that the peak in DOS can serve as a telltale of the guided modes.Parameter values used: $\lambda=-1.5\hbar v$,  energies $\epsilon=\epsilon_0$, $0.5\epsilon_0$, $0.1\epsilon_0$, where $\epsilon_0=\pi\hbar \sqrt{\pi n_0}$, $n_0 = 10^{11}\,\text{cm}^{-2}$ (higher peaks correspond to higher energy $\epsilon$ values).

\newpage

\begin{center}
 \begin{figure}[t]
 \includegraphics[width=0.8\textwidth]{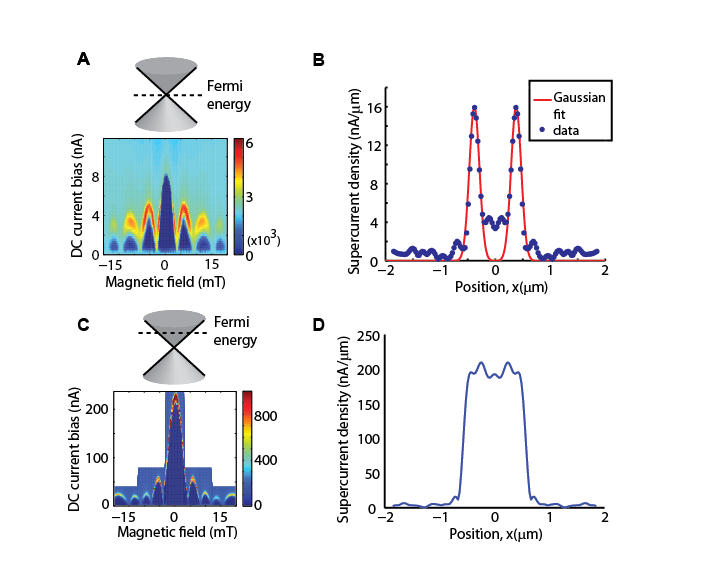}
	  \caption{}
 \end{figure}
\end{center}

\textbf{Fig. S2}.

\textbf{(A)} Edge-dominated 
SQUID-like interference pattern at neutrality in device $ML1$ ($n=2.38\times 10^9$ cm$^{-2}$; colorscale is $dV/dI (\Omega)$).  From Fig. 2A in main text.  \textbf{(B)} Real-space image of current flow confined to the boundaries, from data in part (A). \textbf{(C)} Conventional Fraunhofer pattern for uniform current flow at high electron density ($n=7\times 10^{11}$ cm$^{-2}$).  From Fig. 2E in main text.  \textbf{(D)} Real-space image of current flow confined to the boundaries, from data in part (C).

\newpage

\begin{center}
 \begin{figure}[t]
 \includegraphics[width=1.0\textwidth]{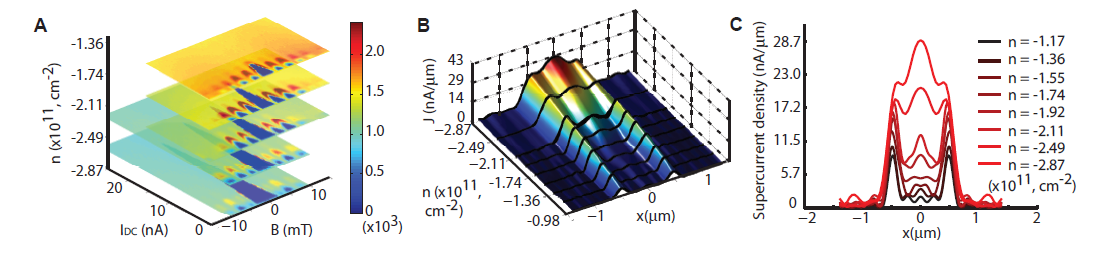}
	  \caption{}
 \end{figure}
\end{center}

\textbf{Fig. S3}.

\textbf{(A)} Sequence of Fraunhofer measurements in bilayer device $BL3$ for the current maps in panels (B) and (C), shown in plots of $dV/dI (\Omega)$ as a function of magnetic field $B$ (mT) and current bias $I_{DC}$ (nA).  \textbf{(B)} Real space image of current flow $J(x)$ as a function of carrier density on the hole side, showing edge currents near the Dirac point and a continuous evolution of bulk flow.  \textbf{(C)} Individual line cuts of $J(x)$ plotted from (B).  This is the data set in Fig. 4A, plotted with a properly scaled vertical axis (supercurrent density, nA/$\mu$m).

\newpage
\begin{center}
\begin{figure*}[htb]
  \includegraphics[width=0.44\textwidth]{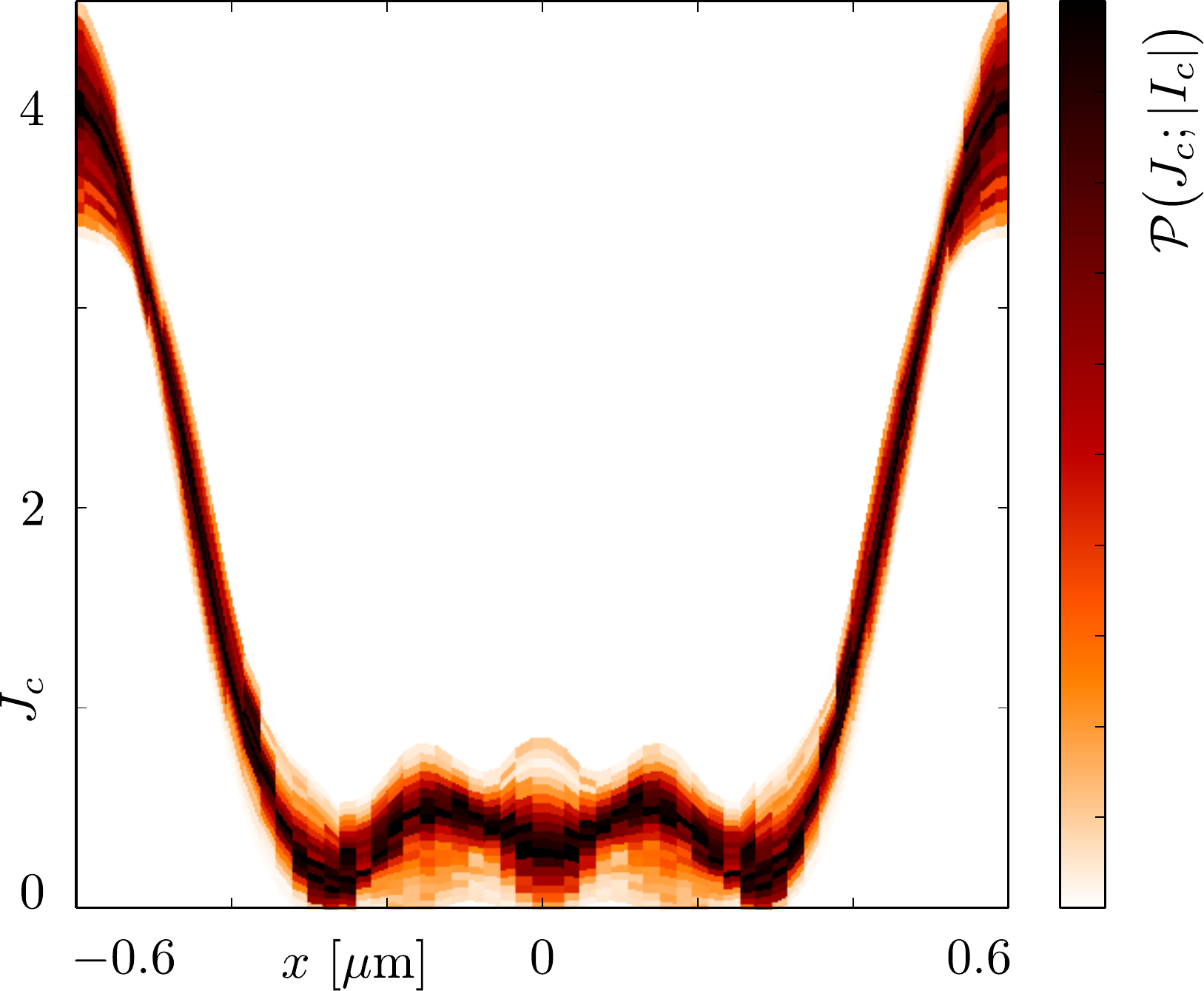}
  \includegraphics[width=0.44\textwidth]{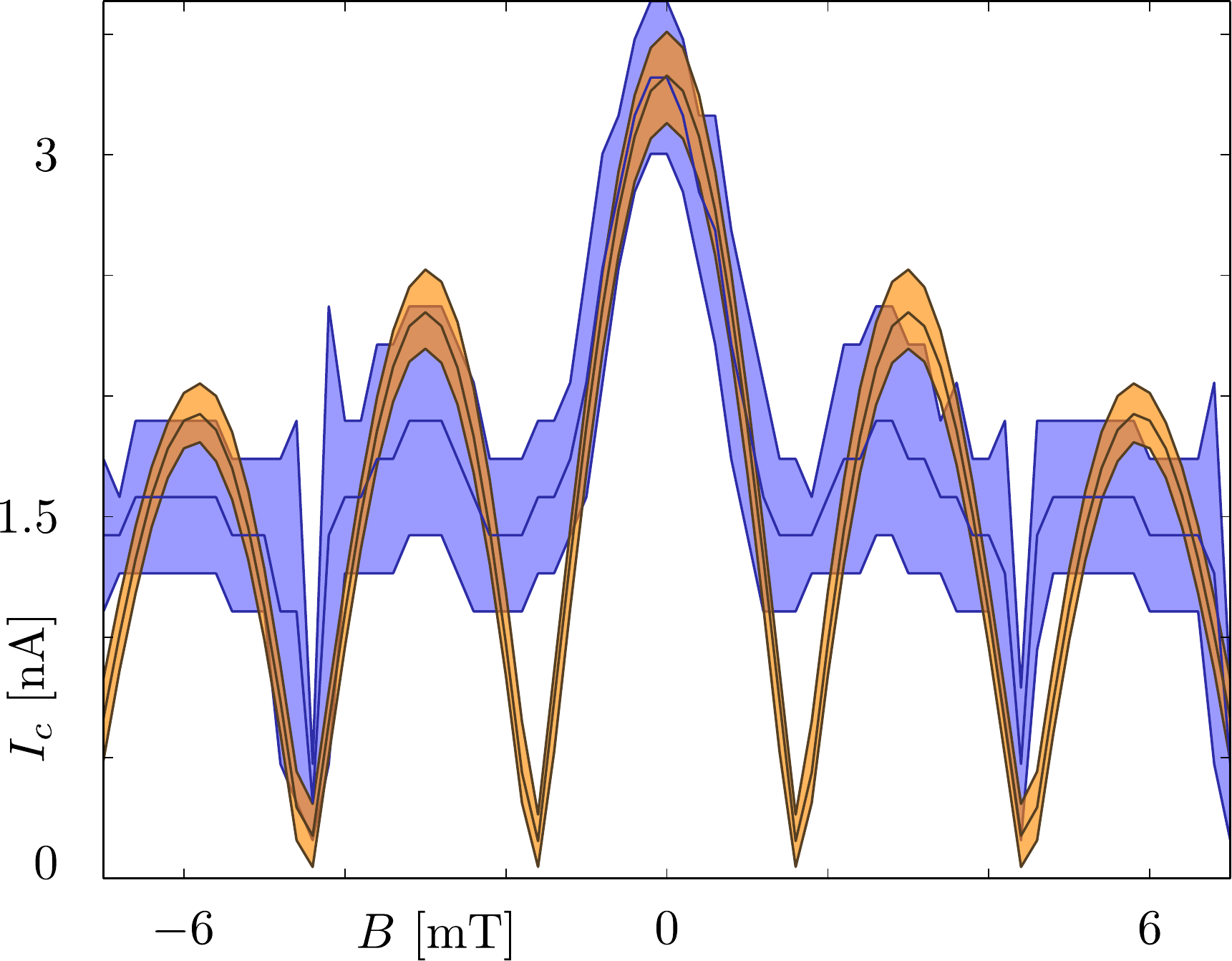}
	  \caption{\label{fig:jcic}}
 \end{figure*}
\end{center}

\textbf{Fig. S4}.

Bayesian estimation of the supercurrent distribution.  Posterior distribution of the current density near the Dirac point in device \textit{BL3} (left panel), and corresponding critical current (right panel). The values of $I_c$ obtained from the posterior distribution (orange) are in good agreement with the measured critical current (blue).

\newpage

\begin{center}
 \begin{figure}[t]
  \includegraphics[width=0.44\textwidth]{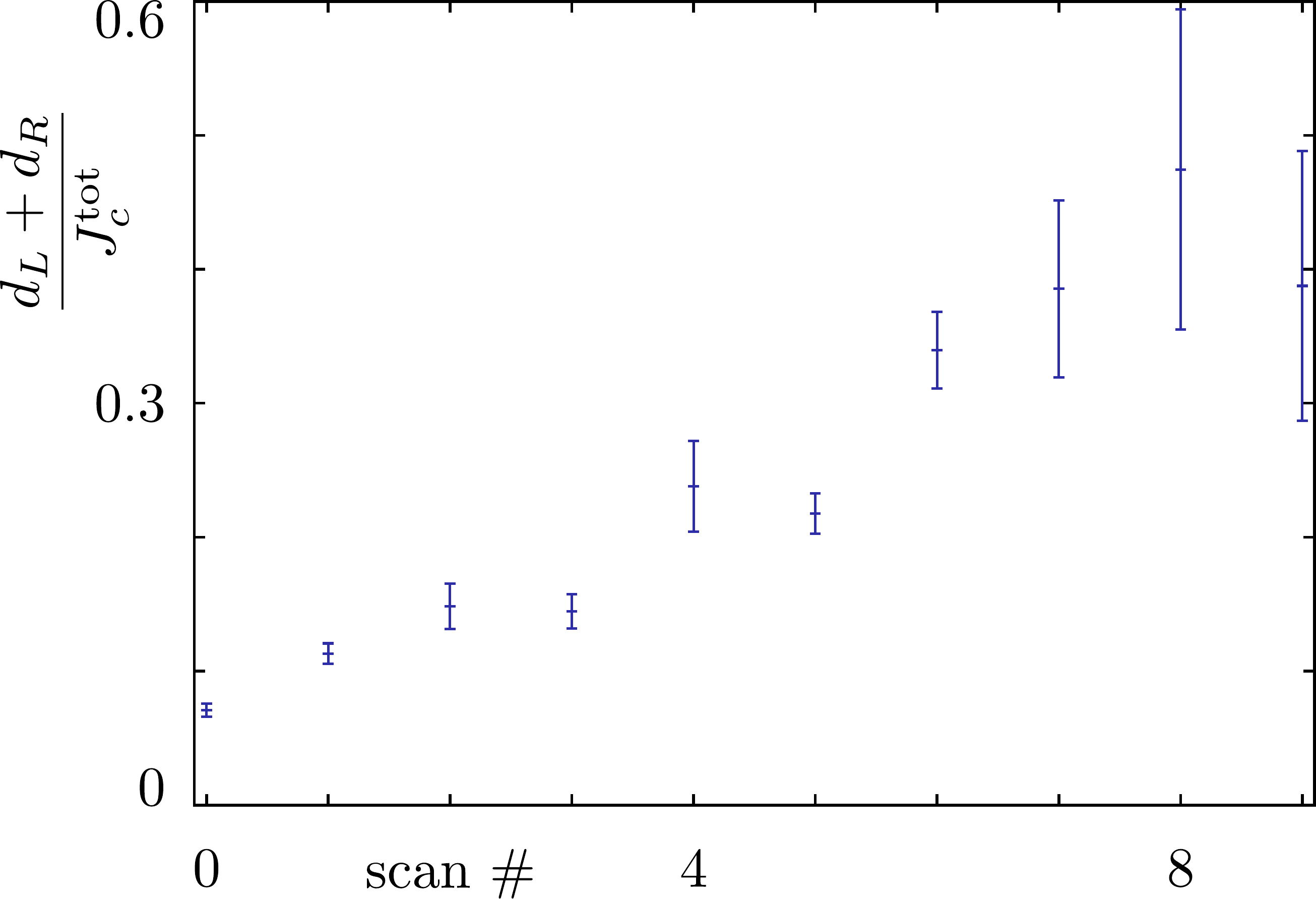}
	  \caption{\label{fig:deltas}}
\end{figure}
\end{center}

\textbf{Fig. S5}.

Ratio of the supercurrent carried by the edge states as a function of carrier density in device \textit{BL3} over the density range of $n\sim -1$ to $-2.9 \times 10^{11}$ cm$^{-2}$. Each scan corresponds to a Fraunhofer pattern, with Fig.~\ref{fig:jcic} showing the 8$^{\rm th}$ scan. (Increasing scan number corresponds to decreasing carrier density.)

\newpage
\textbf{Supplementary Tables}\\

\begin{table}[h]
\begin{tabular}{|c|c|c|c|c|}
\hline
Device  & L (nm) &  W (nm) & Aspect ratio, L/W & Contact width (nm)\\ 
\hline \hline
BL1 & 250 & 1200 & 0.208 & 400\\ 
ML1 & 300 & 1200 & 0.25 & 300 \\ 
BL2 & 300 & 800  & 0.375 & 400 \\ 
BL3 & 350 & 1200 & 0.292 & 600\\ 
BL4 & 250 & 900 & 0.278 & 400\\ 
\hline 
\end{tabular}

\end{table}

\textbf{Table S1}. 

List of device dimensions for the graphene Josephson junctions studied in this work. $L$ and $W$ refer to junction length and width, respectively, as labeled in Fig. 1d of the main text. Contact width refers to the size of the superconducting Ti/Al electrodes in the direction perpendicular to W.  BLx and MLx refer to bilayer and monolayer graphene devices, respectively.

\newpage
\bibliography{bibtex_SC_supp} 